\documentclass[prd,twocolumn,superscriptaddress,floatfix,amsmath,amssymb,amsfonts,longbibliography,nofootinbib]{revtex4-2}

\usepackage{float} 

\usepackage{comment}
\usepackage[normalem]{ulem}
\usepackage[english]{babel}
\usepackage{graphicx}
\usepackage{dcolumn}
\usepackage{bm}
\usepackage{blindtext}
\usepackage{verbatim}
\usepackage{mathrsfs}
\usepackage{musicography}
\usepackage{amsmath}
\usepackage{dsfont}
\usepackage{cancel}
\usepackage{physics}
\usepackage{epstopdf}
\usepackage{mathtools}
\usepackage{color}
\usepackage[usenames,dvipsnames]{pstricks}
\usepackage{epsfig}
\usepackage{pst-grad} 
\usepackage{pst-plot} 
\usepackage{hyperref}
\usepackage{verbatim}
\usepackage{afterpage}
\usepackage{sidecap}

\hypersetup{
    colorlinks=true,
    linkcolor=blue,
    filecolor=red,      
    urlcolor=cyan,
}

\newcommand{\ii}{\mathrm{i}}

\newcommand{\Sp}{{\text{Sp}(2N, \mathbb{R})}}

\begin{document}

\title{Toward a physically motivated notion of Gaussian complexity geometry}

\author{Bruno de S. L. Torres}
\email{bdesouzaleaotorres@perimeterinstitute.ca}
\affiliation{Perimeter Institute for Theoretical Physics, Waterloo, Ontario, N2L 2Y5, Canada}
\affiliation{Department of Physics and Astronomy, University of Waterloo, Waterloo, ON N2L 3G1, Canada}
\affiliation{Institute for Quantum Computing, University of Waterloo, Waterloo, Ontario, N2L 3G1, Canada}

\author{Eduardo Mart\'{i}n-Mart\'{i}nez}
\email{emartinmartinez@uwaterloo.ca}
\affiliation{Department of Applied Mathematics, University of Waterloo, Waterloo, Ontario, N2L 3G1, Canada}
\affiliation{Perimeter Institute for Theoretical Physics, Waterloo, Ontario, N2L 2Y5, Canada}
\affiliation{Department of Physics and Astronomy, University of Waterloo, Waterloo, ON N2L 3G1, Canada}
\affiliation{Institute for Quantum Computing, University of Waterloo, Waterloo, Ontario, N2L 3G1, Canada}

\begin{abstract}

We present a general construction of a geometric notion of circuit complexity for Gaussian states (both bosonic and fermionic) in terms of Riemannian geometry. We lay out general conditions that a Riemannian  metric function on the space of Gaussian states should satisfy in order for it to yield a physically reasonable measure of complexity. This general formalism  can naturally accommodate modifications to complexity geometries that arise from cost functions that depend nontrivially on the instantaneous state and on the direction on circuit space at each point. We explore these modifications and, as a particular case, we show how to account for time-reversal symmetry breaking in measures of complexity, which is often natural from an experimental (and thermodynamical) perspective, but is absent in commonly studied complexity measures. This establishes a first step towards building a quantitative, geometric notion of complexity that faithfully mimics what is experienced as ``easy'' or ``hard'' to implement in a lab from a physically motivated point of view.

\end{abstract}

\maketitle

\section{Introduction}

The notion of complexity quantifies the ``hardness'' of performing a certain task, given some set amount of relevant resources available. The concept originates in computer science and information theory, where it refers to the minimum cost (as measured by resources such as memory usage or number of elementary operations) of a given computation~\cite{BasicComplexity}. This notion is readily exportable to quantum computing~\cite{CleveComplexity, WatrousComplexity}, where a natural measure of complexity is provided by the so-called \emph{circuit complexity}. Given some fixed set of ``simple'' unitary transformations (i.e., an elementary gate set), the circuit complexity of a unitary $\hat{U}$ is the minimum number of elementary gates needed to build $\hat{U}$, and the circuit complexity of a state $\ket{\psi}$ is the minimum number of elementary gates needed to prepare $\ket{\psi}$ starting from some fiducial reference state $\ket{\psi_0}$.

The focus on the cost associated with a given set of transformations is the trademark of complexity, and is not explicitly present in any other information measure. At a practical level, this can be a rather useful tool in characterizing one's ability to perform information-processing tasks. From a more foundational point of view, we now have also come to understand that the physical limitations on how information is processed and manipulated can actually play an important role in fundamental aspects of physical theories, which is why the notion of complexity has appeared as a relevant concept in many areas of physics extending far beyond quantum computing and quantum information.

In many-body physics, complexity plays a central role in understanding chaos and thermalization in closed quantum systems~\cite{ComplexityThermalization1, BeniChaosComplexity, ComplexityThermalization2}, and it also serves as a powerful figure of merit to characterize quantum phases of matter~\cite{QuantumPhasesComplexity1, QuantumPhasesComplexity2, QuantumPhasesComplexity3}. In high energy physics, complexity has also been suggested as an important element in establishing the regimes of validity of quantum field theory in curved spacetimes, something that has led to potential insights into the black hole information loss problem and the nature of the black hole interior in quantum gravity~\cite{HarlowHayden2013, Susskind1a, Susskind1b, Susskind2016, Kim2020, NonIsometricCodes2022}. Finally, the concept of complexity is also of great importance in the context of the AdS/CFT correspondence, where a variety of different geometric quantities of the bulk spacetime have been put forward as potential candidates for the holographic dual of the circuit complexity of the state on the boundary CFT~\cite{ComplexityEqualsVolume, ComplexityEqualsAction, ComplexityEqualsAction2, ComplexityEqualsAction2.0, ComplexityEqualsAnything1, ComplexityEqualsAnything2}.

However, quantitatively characterizing complexity is a very challenging task, since finding the optimal circuit that implements a given unitary or prepares a given state is highly nontrivial in general. In order to tackle this problem, geometric methods of describing complexity have proven to be very powerful. This has been made most concrete by Nielsen's geometric approach~\cite{nielsen2005, nielsen2006, othernielsen2006}, which proposes to reformulate circuit complexity fully in terms of differential geometry. The idea is that by equipping the space of unitary operators with some notion of distance, the complexity of a given unitary can be pictured as the length of a geodesic connecting the identity operator to the unitary of interest. Nielsen's approach has recently been shown to be capable of providing lower bounds to the complexity of typical unitaries which improve upon previously known bounds~\cite{Brown2023}. It has also been very fruitful in establishing concrete notions of complexity in quantum field theory both for free/weakly interacting fields~\cite{Jefferson2017, Hackl2018, Bhattacharyya2018, Chapman2019} and for CFTs~\cite{Caputa2019, Chagnet2022}, which guides much of the recent work in holographic complexity in AdS/CFT (for a recent review, see~\cite{Chapman2022}). Another proposal for the geometrization of complexity is based directly on the geometry on the Hilbert space of quantum states provided by the Fubini-Study metric~\cite{Pastawski2017, Flory2020}. Finally, in a similar spirit, a recent proof of an important conjecture about the growth of complexity in typical random circuits (which was originally motivated by considerations on black holes~\cite{susskind2018, SecondLawComplexity}) was made possible thanks to tools from differential topology and algebraic geometry~\cite{Haferkamp2022}.

Despite all this progress, the inherent ambiguities in key elements of the definition of circuit complexity (namely, in the choices of reference state and elementary gate set, as well as the cost associated to each elementary gate) are sometimes seen as obstacles to a more complete understanding of complexity in QFT and the various related geometrical quantities that are conjectured to be dual to it in AdS/CFT~\cite{Yang2022}. The usual philosophy when dealing with these issues in the context of quantum information theory and quantum computing is to formulate questions about complexity that are less sensitive to these ambiguities. This is the basis, for example, of the definition of quantum complexity classes, which are only concerned with how the complexity of a family of circuits, built with a common set of elementary gates, scales with the size of the input~\cite{WatrousComplexity}. In this paper, however, we will adopt a complementary point of view according to which these seemingly negative aspects of the definition of circuit complexity are actually features of the formalism that can encode relevant physics. After all, if complexity quantifies the difficulty in performing a certain task, it makes sense for it to be determined dynamically, based on what physical operations objectively cost in terms of available lab resources. 

With this motivation in mind, in this paper we will show how to formulate general Riemmanian metrics on the space of pure Gaussian states (both for bosonic and fermionic systems) which are sufficiently broad to accommodate all physically relevant settings, while still having enough structure to allow us to make concrete statements characterizing state complexity. We will also propose an extension to the notion of complexity geometry which allows for a ``non-reversible'' cost---that is, we provide a concrete example of a modified complexity measure which can assign different costs to directions in circuit space related to each other by time reversal. The goal is to yield an operationally motivated measure of complexity that is directly connected to an experimentalist's view of what tasks would be considered easy or hard in practice; this paper provides a step in that direction. The choice of Gaussian states is motivated by the plethora of powerful analytical techniques and results that the assumption of Gaussianity provides, as well as by the pervasiveness of Gaussian states in a variety of different scenarios of physical interest ranging from quantum field theory in curved spacetimes~\cite{Ashtekar1975, Wald2} to quantum optics 
and quantum information with continuous variables more broadly~\cite{ContinuousVariablesQI, gaussianquantuminfo, Adesso2014}. We should note, however, that it is possible to extend this general strategy to other families of circuits whose infinitesimal generators form a Lie algebra. 

The paper is organized as follows:
\begin{itemize}
    \item[$\cdot$] In Sec.~\ref{Sec:ComplexityBackground} we review the main ingredients of Nielsen's geometric approach to circuit complexity~\cite{nielsen2005, nielsen2006, othernielsen2006}, which will be the basis for the formalism employed in the paper. As a preparation for the approach that will be taken later on, we also argue at the end of the Section why adaptations of Nielsen's complexity geometry which relax some of the usual assumptions made about the complexity metric are reasonable and physically motivated. Section~\ref{Sec:GaussianStatesParametrization} presents the unified description of bosonic and fermionic Gaussian states that we will make use of, closely following the formalism of~\cite{Hackl2018, Chapman2019, HacklKahler2021}. While not containing any substantially new material, we believe that the content of Sections~\ref{Sec:ComplexityBackground} and~\ref{Sec:GaussianStatesParametrization} is useful for fixing notation and terminology, and also helps to contextualize the results to be presented later in the paper with what is already known in the literature.
    \item[$\cdot$] Section~\ref{Sec:gengaussianmetric} describes the general conceptualization of a Riemannian metric that can yield a notion of complexity geometry for pure Gaussian states.  
The main novel contribution of this section is the formulation of a minimal, well-motivated assumption that the metric will be required to satisfy, under which the state complexity defined by any physically reasonable metric on the space of Gaussian states can be characterized in a unified way. This condition imposed on the (otherwise quite general) metric on circuit space is well-motivated by the preferred choice of reference state relative to which we define the complexity of the target state. This also extends known results and methods in the literature to a much more general set of metrics, with potentially arbitrary cost functions assigned to a given elementary gate set.
\item[$\cdot$] In Section~\ref{sec:notimereversal} we propose a new extension to the complexity geometry which introduces a feature of ``non-reversibility'' to the complexity measure by assigning different costs to a given gate and its inverse. This is another arguably very physical feature of an experimentalist's experience of what is easy or hard to be achieved in a lab, which is however usually absent in most adopted measures of complexity geometry. The particular new ingredient proposed has a very intuitive physical interpretation as a kind of vector potential, which modifies the equation of motion for the optimal paths in a way that directly reproduces the Lorentz force law experienced by a charged particle in a background electromagnetic field.
\item[$\cdot$] Finally, Section~\ref{Sec:Applications} illustrates the general strategy with a few simple examples where state complexity can actually be computed in closed form, and Section~\ref{ConclusionSection} summarizes our main results and discusses potential future directions.
\end{itemize}
  
\section{Nielsen's geometric approach to circuit complexity}\label{Sec:ComplexityBackground}

Let $\mathcal{U}$ be the set of all unitary operators that act on some Hilbert space $\mathcal{H}$ describing the physical system of interest. The first step in Nielsen's geometric approach to circuit complexity is to replace a discrete quantum circuit by a continuous path $\hat{U}(t)$ in $\mathcal{U}$, which can in general be written as
\begin{equation}\label{eq:timeevolutionop}
    \hat{U}(t) = \mathcal{T}\exp\left(-\ii \int^t_{t_0} \hat{H}(t')\dd t'\right).
\end{equation}
The (in general time-dependent) operator $\hat{H}(t)$ (let us call it `Hamiltonian' for short)  can essentially be seen as equivalent to the tangent vector to the trajectory $\hat{U}(t)$, as Eq.~\eqref{eq:timeevolutionop} is a solution to the differential equation
\begin{equation}\label{eq:schrodinger}
    \dfrac{\dd \hat{U}}{\dd t} = -\ii \hat{H}(t)\hat{U}(t),
\end{equation}
which defines the relation between the tangent vector $\dd \hat{U}/\dd t$ and the Hamiltonian generator $\hat{H}(t)$.

Nielsen's geometric approach then replaces the notion of a ``number of gates'' applied at each layer of the circuit by a \emph{cost function} $c(\hat{U}, \hat{V})$ which assigns to each infinitesimal time step $\dd t$ taken at a point $\hat{U}$ with tangent vector $\dd \hat{U}/\dd t = \hat{V}$ an infinitesimal cost
\begin{equation}\label{eq:infcost}
    \dd C = c(\hat{U}(t), \hat{V}(t))\,\dd t,
\end{equation}
where the Hamiltonian $\hat{H}(t)$ that generates this step can be given in terms of the tangent vector $\hat{V}$ by inverting Eq.~\eqref{eq:timeevolutionop},
\begin{equation}\label{eq:HamiltonianFromTangentVector}
    \hat{H}(t) = \ii \hat{V}(t)\hat{U}^\dagger(t).
\end{equation}
The finite version of the cost of a given circuit is thus given by
\begin{equation}\label{eq:finitecost}
    C= \int_{t_0}^{t_f} c(\hat{U}(t), \hat{V}(t))\,\dd t
\end{equation}
and the complexity of a unitary $\hat{U}_T$ is defined as
\begin{equation}\label{eq:geometriccomplexity}
    \mathcal{C}(\hat{U}_T) = \min\left(\int_{t_0}^{t_f} c(\hat{U}(t), \hat{V}(t))\,\dd t\right)
\end{equation}
where the minimization ranges over all the trajectories $\hat{U}(t)$ in $\mathcal{U}$ satisfying the boundary conditions
\begin{equation}
    \hat{U}(t_0) = \hat{\mathds{1}}, \,\, \hat{U}(t_f) = \hat{U}_T.
\end{equation}
Similarly, one defines the state complexity of a target state $\ket{\psi_T}$ relative to a reference state $\ket{\psi_R}$ as
\begin{equation}
    \mathcal{C}(\ket{\psi_R}, \ket{\psi_T}) = \min_{\hat{U}_T}\{\mathcal{C}(\hat{U}_T)\,\,|\,\,\hat{U}_T\ket{\psi_R} = \ket{\psi_T}\}.
\end{equation}

The most basic requirement that we would expect from $c(\hat{U}, \hat{V})$ is that it should be a smooth and positive-definite function---that is, for any unitary $\hat{U}$ and any tangent vector $\hat{V}$ in the tangent space to $\hat{U}$, it holds that $c(\hat{U}, \hat{V}) \geq 0$. It is usual to demand that the inequality be saturated if and only if $\hat{V} = 0$. After all, it is natural to expect the cost of any given infinitesimal step to be nonnegative, and only be zero if no gates are applied at all. It also makes sense to require that $c(\hat{U}, \hat{V})$ satisfy the triangle inequality in the second argument, $c(\hat{U}, \hat{V}_1 
+ \hat{V}_2) \leq c(\hat{U}, \hat{V}_1) + c(\hat{U}, \hat{V}_2)$. Intuitively, this just tells us that we cannot decrease the cost of a given step by decomposing the tangent vector at that point into a given linear combination, and then evolving with each component of the linear combination separately.  Finally, if we want the depth given by Eq.~\eqref{eq:finitecost} to be invariant under general reparametrizations $t \to f(t)$ that preserve the time orientation of the path (that is, $\dd f/\dd t > 0$), we would also require that $c$ be positive-homogeneous of degree $1$ in the second argument,\footnote{We only demand positive-homogeneity (instead of requiring, for instance, that $c(\hat{U}, \lambda\hat{V}) = \abs{\lambda} c(\hat{U}, \hat{V})$ for every $\lambda \in \mathbb{R}$) because we would also like to accommodate cases where the cost is not invariant under time reversal. Although most of the examples studied in practice do assign the same cost to a given elementary gate and its inverse, in Sec.~\ref{sec:notimereversal} we give an example where that is not the case.} i.e., $c(\hat{U}, \lambda\hat{V}) = \lambda c(\hat{U}, \hat{V})$ for every $\lambda \geq 0$. Under the assumption that Eq.~\eqref{eq:finitecost} is reparametrization-invariant, we can always choose the parameter $t$ to be such that the initial point of the curve $\hat{U}(t)$ is at $t_0=0$ and the final point is at $t_f = 1$. That is the convention that we will adopt for the rest of the paper.

The conditions we just provided give $c(\hat{U}, \hat{V})$  the properties of a \emph{Finsler metric}, which then equips the unitary group with the structure of a \emph{Finsler manifold}. Finsler manifolds are a generalization of Riemannian manifolds where one can still talk about lengths between points, but the length functional is not necessarily derived from a metric tensor. The interpretation of the cost function as a Finsler metric allows us to see expression~\eqref{eq:finitecost} as the length of the curve $\hat{U}(t)$ between times $t_0$ and $t_f$, and the optimal circuit that prepares a given unitary $\hat{U}_T$ is nothing more than the trajectory of minimum length---i.e., the geodesic---connecting $\hat{\mathds{1}}$ and $\hat{U}_T$. The problem of computing circuit complexity then becomes a problem in variational calculus, where the full power of differential geometry can be put in action.

In order to put this (somewhat abstract) notion of a Finsler metric on the unitary group in closer contact with the original idea of circuit complexity, it is customary to expand the operator $\hat{H}(t)$ generating the circuit at each instant of time in a fixed basis of Hermitian generators $\hat{\mathcal{O}}_I$,
\begin{equation}\label{eq:hamiltonianexpansion}
    \hat{H}(t) = \sum_I Y^{I}(t)\hat{\mathcal{O}}_I,
\end{equation}
and express the cost function explicitly in terms of the components $Y^I$ in this basis. The interpretation of the generators $\hat{\mathcal{O}}_I$ is that the unitaries $\hat{U}_I = e^{\ii \varepsilon \hat{\mathcal{O}}_I}$ form our elementary gate set, with $\varepsilon$ being some minimum time scale for each gate to be applied. 

If our complexity measure only cares about the number of gates applied at each time step, the cost function evaluated at a given time $t$ along the path in $\mathcal{U}$ will only depend on the Hamiltonian $\hat{H}(t)$ generating the circuit at that time. In more technical terms, this means that $c(\hat{U}, \hat{V})$ defines a \emph{right-invariant} Finsler metric, which satisfies
\begin{equation}\label{eq:rightinvarianceFinsler}
    c(\hat{U}, \hat{V}) = c(\hat{\mathds{1}}, \hat{V}\hat{U}^\dagger).
\end{equation}
If we parametrize the tangent vector $\hat{V}$ in terms of the Hamiltonian $\hat{H}(t)$ through Eq.~\eqref{eq:HamiltonianFromTangentVector} and we expand $\hat{H}(t)$ in a given basis of generators as in~\eqref{eq:hamiltonianexpansion}, right-invariance of the Finsler metric as in Eq.~\eqref{eq:rightinvarianceFinsler} implies that the cost function only depends on the components $Y^I$, and not on the point $\hat{U}$; in short, $c(\hat{U}, \hat{V})\equiv F(Y)$ for some function $F$. 
Simple examples of cost functions $F(Y)$ commonly considered in the literature include~\cite{nielsen2005}
\begin{subequations}
    \begin{align}
        F_1(Y) &= \sum_I \abs{Y^I}, \label{eq:F1metric}\\
          F_{1p}(Y) &= \sum_I p_I\abs{Y^I}, \label{eq:F1pmetric}\\
        F_2(Y) &= \sqrt{\sum_I \left(Y^I\right)^2}, \label{eq:F2metric}\\
        F_{2q}(Y) &= \sqrt{\sum_I q_I\left(Y^I\right)^2}. \label{eq:F2qmetric}
    \end{align}
\end{subequations}
The cost function~\eqref{eq:F1metric} (unoriginally named the $F_1$\emph{-metric}) is the one that most closely matches the original idea of counting the number of gates in the circuit. The $F_{1p}$-metric~\eqref{eq:F1pmetric} is a modification of~\eqref{eq:F1metric} that includes possibly nontrivial \emph{penalty factors} $p_I$ that may vary between different gates. This can be used to enforce further physical constraints in the complexity measure, for instance, by making nonlocal gates more costly than local ones. Strictly speaking, the $F_1$-metric is not Finsler, since it violates the smoothness requirement; as such, it is not immediately amenable to the methods from variational calculus and differential geometry. As already explained in e.g.~\cite{nielsen2005}, this can be remedied by approximating $F_1$ by a one-parameter family of Finsler metrics that recover the $F_1$ metric in some limit. Another (more common) approach is to instead adopt the $F_2$-metric~\eqref{eq:F2metric}, which has the natural interpretation of a \emph{Riemannian} metric that emerges from an Euclidean inner product on the tangent space at each point of $\mathcal{U}$. This is the strategy most explored in applications of Nielsen's framework to complexity in quantum field theory~\cite{Jefferson2017, Hackl2018, Chapman2019}. Naturally, the $F_{2q}$-metric~\eqref{eq:F2qmetric} is to $F_2$ the same as $F_{1p}$ is to $F_1$.

In this paper, however, we are interested in a generalization of geometric measures of complexity where the cost function depends nontrivially on the point in circuit space where it is being evaluated. In this case, the Finsler metric is no longer right-invariant, and therefore the measure of complexity at hand is more detailed than a simple counting of the number of gates in a circuit. Physically, this is very well-motivated: after all, it is possible to consider regimes where, from the point of view of an experimentalist, the ability to reliably apply a given infinitesimal transformation will depend on the state of the system. One can imagine, for instance, that increasing the entanglement between two parties by a given amount may become harder if the state you are acting on is already highly entangled, as compared to when the two parties start in a product state. 

Additionally, previous geometric complexity measure proposals do not address the issue of non-reversibility of complexity. For example, acting on the ground state of a system to prepare some complex entanglement structure is much more experimentally involved than starting from some very entangled state and reaching the ground state, since for the latter it just becomes a matter of waiting for the system to decay to its ground state when coupled to some cold enough environment. In terms of the actual lab resources (e.g., money, graduate students, etc.) spent in the experiment, the second scenario is arguably cheaper than the first, and it would be relevant to have a measure of complexity that mimics this feature. 

As we will see shortly, several important features of the geometry of the optimal circuit can be stated in a way that does not actually depend on right-invariance, so it is possible to extend known results and approaches for right-invariant metrics to more general Riemannian metrics on the space of unitaries. We will also see how to modify the cost functional in order to incorporate physical features of irreversibility in the complexity measure in a very natural geometrical way. This will establish the basis on which to build a formalism for the systematic study of complexity measures that are, in some sense, closer to an experimentalist's view on what is easy or hard to perform. 
 
\section{Parametrization of pure Gaussian states}\label{Sec:GaussianStatesParametrization}
Our analysis in this paper will focus a general description of complexity geometry for a special class of states known as \emph{Gaussian states}. These can be generally interpreted as squeezed, coherent, thermal states of systems with quadratic Hamiltonians. Thanks to the fact that the Hamiltonian governing the dynamics of every system close to equilibrium can be approximated as quadratic, Gaussian states are ubiquitous in the study of a wide range of systems in a variety of areas of physics. Applications range from foundational aspects of quantum field theory in curved spacetimes~\cite{Ashtekar1975, Wald2} to experimental implementations of quantum information protocols with quantum optics~\cite{Gottesman2001, KLM}. The mathematical formalism for dealing with Gaussian states for both bosonic and fermionic systems is also very well-developed, and forms the basis for much of the results in quantum information and many-body physics with continuous variables~\cite{ContinuousVariablesQI, gaussianquantuminfo, ScramblingPhaseSpace}.

We will closely follow a strategy employed in~\cite{Hackl2018, Chapman2019} which parametrizes pure Gaussian states as equivalence classes of an appropriately chosen matrix group describing the space of unitaries of interest. As we will see shortly, this approach is very powerful and natural for considerations about Gaussian state complexity. 

Consider a quantum system with $N$ degrees of freedom (or modes), described by $2N$ quadrature operators $\hat{\xi}^a$. These consist of a set of Hermitian operators,
\begin{equation}
	\hat{\xi}^a = (\hat{Q}^1, \hat{P}_1, \dots, \hat{Q}^N, \hat{P}_N)^\intercal
\end{equation}
which, in the case of bosons, satisfy the canonical commutation relations
\begin{equation}\label{eq:CCR}
	\begin{gathered}
		[\hat{Q}^i, \hat{P}_j] = \ii \delta^{i}_j\hat{\mathds{1}}, \\
		[\hat{Q}^i, \hat{Q}^j] = [\hat{P}_i, \hat{P}_j] = 0,
	\end{gathered}
\end{equation}
or, in the case of  fermions, the \emph{anti}commutation relations 
\begin{equation}
	\begin{gathered}
	\{\hat{Q}^i, \hat{P}_j\} = 0, \\
	\{\hat{Q}^i, \hat{Q}^j\} = \{\hat{P}_i, \hat{P}_j\} = \delta^i_j\hat{\mathds{1}}.
\end{gathered}
\end{equation}
Referring to the phase-space operators $\hat{\xi}^a$ as ``quadratures'' is standard in the literature for bosonic systems; for fermionic systems, one might find it more usual to call this basis of Hermitian generators ``Majorana modes''. However, since most of what we will present in what follows will apply both to fermionic and bosonic systems in a unified fashion, we will use ``quadratures'' to refer to both cases at once, in line with the conventions used, e.g., in~\cite{Windt2021}. 

A Gaussian state $\hat{\rho}$ is fully characterized by the one-point and two-point correlators of the quadratures,
\begin{align}
    z^a&\coloneqq  \langle\hat{\xi}^a\rangle_{\hat{\rho}},\\
    W^{ab} &\coloneqq \langle\hat{\xi}^a\hat{\xi}^b\rangle_{\hat{\rho}} = \dfrac{1}{2}\left(\sigma^{ab} + \ii \Omega^{ab}\right) + z^a z^b,\label{eq:twopointfunction}
\end{align}
where we have defined
\begin{align}
    \sigma^{ab} &= \langle\hat{\xi}^a\hat{\xi}^b + \hat{\xi}^b\hat{\xi}^a\rangle_{\hat{\rho}} - 2 z^a z^b, \\
    \Omega^{ab} &= -\ii\langle\hat{\xi}^a\hat{\xi}^b - \hat{\xi}^b\hat{\xi}^a\rangle_{\hat{\rho}}.\label{eq:sympmatrix}
\end{align}
Higher-order $n$-point correlators  are then directly obtained from these one- and two-point correlators~\eqref{eq:twopointfunction} via Wick contractions. Note that for fermions, the one-point function always vanishes.\footnote{We should note that there is such a thing as coherent states for fermionic systems, where one can have a nonzero displacement $z^a$ as a Grassman-valued vector. While these can be useful calculational tools, there are no actually physical fermionic Gaussian states~\cite{HacklKahler2021}.  Therefore, we will only assume nonzero one-point functions the description of bosonic Gaussian transformations, as it is common in phase space quantum mechanics.} In this case, the formalism simplifies, as we can restrict ourselves to the two-point function only. 

Considered as tensors on phase space, the real part $\sigma^{ab}$ of the two-point correlator defines  a metric (in the sense that it is a symmetric, positive-definite bilinear map), and the  imaginary antisymmetric part $\Omega^{ab}$ defines  a symplectic form (in the sense that it is a antisymmetric and non-degenerate bilinear map). Due to the commutation (anticommutation) relations of the quadratures in bosonic (fermionic) systems, only the real (imaginary) part of the two-point function actually carries nontrivial information about the correlations in the Gaussian state, with the remaining part being fully state-independent and fixed by the algebra of the quadratures. It is common to refer to the state-dependent part of the correlations in a Gaussian state ($\sigma^{ab}$ for bosons, and $\Omega^{ab}$ for fermions) as the \emph{covariance matrix}. We can then combine the state-dependent and state-independent parts of the correlations in a Gaussian state into a single phase-space operator which we call the \emph{complex structure},
\begin{equation}\label{eq:defComplexStructure}
    J^{a}_{\phantom{a}b} \coloneqq \begin{dcases*}-\sigma^{ac}\Omega_{cb} & for bosons \\
    \Omega^{ac}\sigma_{cb} & for fermions,
    \end{dcases*}
\end{equation}
where we have defined $\sigma_{ab}$ and $\Omega_{ab}$ as the matrix inverses of $\sigma^{ab}$ and $\Omega^{ab}$ respectively\footnote{For mixed fermionic Gaussian states, it might happen that $\Omega^{ab}$ is not invertible. In this case, $\Omega_{ab}$ is defined as the pseudoinverse of $\Omega^{ab}$ with respect to $\sigma^{ab}$---i.e., we invert $\Omega^{ab}$ only on the subspace that is orthogonal (in the sense of the inner product defined by $\sigma^{ab}$) to its kernel~\cite{Windt2021, Hackl2020}.},
\begin{align}
    \sigma_{ab} &\coloneqq \left(\sigma^{-1}\right)_{ab}, \\
    \Omega_{ab} &\coloneqq \left(\Omega^{-1}\right)_{ab}.
\end{align}
For pure Gaussian states---which will be our main focus in this paper---it turns out that the two (in principle different) definitions for $J^{a}_{\phantom{a}b}$ in~\eqref{eq:defComplexStructure} actually coincide: namely, for both fermions and bosons, $\sigma^{ac}\Omega_{cb} = -\Omega^{ac}\sigma_{cb}$ if the state is pure. In this case, the operator defined in~\eqref{eq:defComplexStructure} satisfies $J^2 = -\openone$ (or $J^{a}_{\phantom{a}c}J^c_{\phantom{c}b} = -\delta^a_b$ in index notation), which justifies calling it a complex structure. For mixed states the equality no longer holds, but it is somewhat customary to keep the name regardless. 

Unitary dynamics preserving Gaussianity (in short, unitary Gaussian transformations) are generated by Hamiltonians that are at most quadratic in the phase-space variables $\hat{\xi}^a$, and act as affine operators on phase space. One can thus view Gaussian dynamics as implemented by pairs $(\bm{v}, M)$, where $v \in \mathbb{R}^{2N}$ is a phase-space displacement vector, and $M \in \text{GL}(2N, \mathbb{R})$ is some invertible matrix. A Gaussian transformation labeled by $(\bm{v}, M)$ will act on the quadratures (in the Heisenberg picture) as
\begin{equation}
    \hat{\xi}^a \mapsto M^{a}_{\phantom{a}b}\hat{\xi}^b + v^a\hat{\mathds{1}}.
\end{equation}
The vector of means therefore evolves as
\begin{equation}
    z^a \mapsto M^{a}_{\phantom{a}b}z^b + v^a,
\end{equation}
and the matrices $\sigma^{ab}$ and $\Omega^{ab}$ evolve as
\begin{align}
    \sigma^{ab} \mapsto M^{a}_{\phantom{a}c}\sigma^{cd}M^{b}_{\phantom{b}d}, \\
    \Omega^{ab} \mapsto M^{a}_{\phantom{a}c}\Omega^{cd}M^{b}_{\phantom{b}d},
\end{align}
or, in matrix notation,
\begin{align}
    \bm z \mapsto M\bm z + \bm v,\\
    \sigma \mapsto M \sigma M^{\intercal}, \\
    \Omega \mapsto M \Omega M^{\intercal}
\end{align}
where $M^\intercal$ is the transpose of $M$, with matrix elements $\left(M^\intercal\right)_{b}^{\phantom{b}a} = M^a_{\phantom{a}b}$. Finally, the complex structure $J$ transforms as
\begin{equation}
    J \mapsto M J M^{-1}.
\end{equation}

Since the canonical commutation/anticommutation relations between the quadratures are state-independent, the matrices $M$ that implement unitary Gaussian transformations must preserve either the symplectic form (in the case of bosons) or the metric given by $\sigma$ (in the case of fermions). Gaussian transformations for systems of bosons are therefore represented on phase space by pairs $(\bm{v}, M)$ where $v$ is a general vector in $\mathbb{R}^{2N}$ and $M$ is an element of the \emph{symplectic group},
\begin{equation}\label{eq:sympgroup}
    \Sp = \{M \in \text{GL}(2N, \mathbb{R})\,\,|\,\, M\Omega M^\intercal = \Omega\}.
\end{equation}
The set of pairs $(\bm{v}, M) \in \mathbb{R}^{2N}\rtimes \Sp $ forms a new Lie group, with group multiplication being given by
\begin{equation}
    (\bm{v}_1, M_1) \cdot (\bm{v}_2, M_2) = (M_1\bm{v}_2 + \bm{v}_1, M_1 M_2).
\end{equation}
For fermions, there are no displacements, since the one-point function of the quadratures is taken to always vanish. Therefore, fermionic Gaussian transformations are elements of the \emph{orthogonal group} with respect to the inner product defined by the symmetric, positive-definite matrix $\sigma$,
\begin{equation}\label{eq:orthogroup}
    \text{O}(2N) = \{M \in \text{GL}(2N, \mathbb{R})\,\,|\,\, M\sigma M^\intercal = \sigma\}.
\end{equation}

The set of all unitary Gaussian transformations acting in a system of $2N$ quadratures thus forms a finite-dimensional group $G$, both for bosons and for fermions. In the case of bosons, the group is given by \mbox{$G = \mathbb{R}^{2N}\rtimes \Sp$}. For fermions, since we are mostly interested in the subset of orthogonal transformations that can be continuously connected to the identity, we will take $G = \text{SO}(2N)$ instead of the full orthogonal group. If our elementary generators $\{\hat{\mathcal{O}}_I\}$ from~\eqref{eq:hamiltonianexpansion} only contains operators that are at most quadratic in $\hat{\xi}^a$, then the space of available unitary transformations acting on the Hilbert space of either bosons or fermions can be identified, in a one-to-one fashion up to a global phase, with the group $G$, and the at-most quadratic generators $\hat{\mathcal{O}}_I$ can be mapped to elements of the Lie algebra $\mathfrak{g}$ of $G$. This renders the description of trajectories on the space of Gaussian transformations effectively finite-dimensional, since $G$ is a finite-dimensional group. The simplifications that arise from this fact represent the main reason behind the power of the analytical treatments of Gaussian states---especially in the case of bosons, whose unrestricted Hilbert space is infinite-dimensional. 

Now that we have mapped Gaussian unitary transormations to elements of a finite dimensional group $G$, let us briefly turn our attention to states. Bosonic Gaussian states can be identified with pairs $(J, \bm{z})$, where $J$ is a complex structure and $\bm{z}$ is the vector of means of the quadratures. In the case of fermions, the vector of means is always taken to be zero, so the complex structure $J$ alone already encodes all the information about the fermionic Gaussian state. In either case, for any two pure Gaussian states---labeled by $(J_1, \bm{z}_1)$ and $(J_2, \bm{z}_2)$ in the case of bosons, or simply $J_1$ and $J_2$ for fermions---it is always possible to find a pair $(\bm{v}, M) \in \mathbb{R}^{2N}\rtimes \Sp $ (for bosons) or an orthogonal matrix $M\in \text{O}(2N)$ (for fermions) such that
\begin{align}
    (J_2, \bm{z}_2) = (M J_1 M^{-1}, M\bm{z}_1 + v) \,\,\,\,\,\text{(bosons)}&, \\
    J_2 = M J_1 M^{-1} \,\,\,\,\,\,\,\,\,\,\,\,\,\,\,\,\,\,\,\,\text{(fermions)}&.
\end{align}
For the case of fermions, if we restrict $G$ to only include elements of $\text{SO}(2N)$ instead of $\text{O}(2N)$, then a Gaussian transformation connecting two fermionic Gaussian states exists as long as both states belong to the same parity sector---i.e., both have the same fermion parity number~\cite{Hackl2018}.
However, in all cases, fixing the initial and final states does not uniquely fix the Gaussian transformation mapping one to the other. In particular, if a given matrix $M$ (in the symplectic group for bosons, or in the orthogonal group for fermions) is such that $M J_1 M^{-1} = J_2$, then so is $M S$, whenever the matrix $S$ satisfies
\begin{equation}\label{eq:stabilizerU}
    SJ_1 S^{-1} = J_1.
\end{equation}
Therefore, if the initial and final states have vanishing vector of means, the transformation mapping between the two is only well-defined modulo the right action of matrices $M$ which commute with the complex structure of the initial state. 

For a given complex structure $J$, the set of all $S$ satisfying~\eqref{eq:stabilizerU} defines a subgroup of the set of Gaussian transformations which we call the \emph{stabilizer} of $J$, denoted by $\text{Sta}_J(N)$. More precisely, we define
\begin{equation}\label{eq:defStab}
    \text{Sta}_J(N) \coloneqq \begin{dcases*}
        \{S \in \text{SO}(2N),\,\, [S, J] = 0 \}, \\
        \{(\bm{0}, S)\,\,|\,\, S \in \Sp,\,\, [S, J] = 0 \},
    \end{dcases*} 
\end{equation}
where the first line above applies to fermions and the second line applies to bosons. 

In words, $\text{Sta}_J(N)$ simply corresponds to the subset of Gaussian transformations represented by linear operators on phase space that leave the covariance matrix of the system invariant. Since the covariance matrix defines a metric in the case of bosons and a symplectic form in the case of fermions, we conclude that, for any choice of $J$ used as reference, the subgroup $\text{Sta}_J(N)$ is isomorphic to the intersection $\text{Sp}(2N, \mathbb{R})\cap \text{SO}(2N)$. For a Gaussian state described by the complex structure $J$ and vanishing vector of means, $\text{Sta}_J(N)$ simply consists of the Gaussian transformations that do not change the physical state. 

Note that for the case of bosonic Gaussian states with $\bm{z}\neq 0$, the action of $\text{Sta}_J(N)$ \emph{will} change the physical state, as it will in general change the vector of means by $\bm{z} \to S\bm{z}$. We could have defined the stabilizer as including a nonzero displacement that undoes this shift in $\bm{z}$; however, we believe that the choice of stabilizer not containing any displacements is more physically motivated. The intuitive idea is that we want the action of the stabilizer to represent some kind of ``free evolution'' that can act on the reference state at no cost, as long as the structure of correlations present in the reference state is preserved. If we picture this ``free evolution'' as being generated by some Hamiltonian, we consider that having only quadratic terms in the Hamiltonian (i.e., harmonic couplings between the quadratures) is more natural than including both quadratic and linear terms (which we would normally picture as coming from some additional driving force on the quadratures). The subgroup generated only by a family of purely quadratic Hamiltonians, \emph{with no linear terms}, that preserves the structure of correlations in the initial state, is precisely the definition of $\text{Sta}_J(N)$ as written in~\eqref{eq:defStab}. It is worth noting, however, that the formal aspects of the framework that we are about to describe would remain largely unaltered if we had chosen to define the stabilizer as the subset of bosonic Gaussian transformations given by pairs of the form $((\mathds{1}- S)\bm{z}, S)$, where $[S, J] = 0$.

We thus have the following parametrization of the space of pure Gaussian states. First, without loss of generality and for simplicity, we take one zero-mean state $\ket{J_R}$ as a reference (so that the elements of the stabilizer leave it invariant). Then, we note that all pure Gaussian states can be obtained from $\ket{J_R}$ through the action of an element of $G$, but two elements $M_1, M_2 \in G$ characterize the same state if they differ by the right-action of some element of $\text{Sta}_{J_R}(N)$. Therefore, pure Gaussian states can be described as \emph{equivalence classes} of Gaussian transformations, where right-action by an element of $\text{Sta}_{J_R}(N)$ defines the equivalence relation. This means that, once we pick a particular state $\ket{J_R}$ as a reference, we can endow the group $G$ of Gaussian transformations with the structure of a fibre bundle, where the fibres are orbits of the stabilizer of the reference state, and the base manifold---which is isomorphic to the space of pure Gaussian states---is the quotient $G/\text{Sta}_{J_R}(N)$. This is the setting that we will use to parametrize the space of Gaussian states, and eventually characterize state complexity, in the following section. 

\section{General features of Gaussian state complexity geometry}\label{Sec:gengaussianmetric}

So far, none of what we have said depends on a metric on the group 
$G$. The stabilizer subgroup, and therefore the characterization of Gaussian states as equivalence classes of Gaussian transformations, are  functions of only the reference state. The next step in order to define a geometric notion of complexity is to equip $G$ with notions of lengths and angles so that we can turn the question of circuit complexity into a problem in Riemannian geometry.

Let $g$ be a Riemannian metric on $G$. This is a tensor field which assigns, at each point $p \in G$, a bilinear map 
\begin{equation}
\begin{gathered}
    g_p: \mathcal{T}_pG\times \mathcal{T}_pG \rightarrow \mathbb{R} \\
    (V_p, W_p) \mapsto g_p(V_p, W_p) 
    \end{gathered}
\end{equation}
that is symmetric and positive-definite,
\begin{equation}
\begin{gathered}
    g_p(V_p, W_p) = g_p(W_p, V_p)\,\, \forall\,\, V_p, W_p \in \mathcal{T}_pG, \\
    g_p(V_p,V_p)>0 \,\,\forall\,\, V_p \neq 0.
    \end{gathered}
\end{equation}
The norm of a vector with respect to this metric is then just given by the usual expression, $\norm{V_p} = \sqrt{g_p(V_p, V_p)}$.\footnote{We emphasize that this Riemannian metric $g$ is completely different from the metric on phase space that could be defined, for instance, via the symmetric, positive-definite matrix $\sigma$ associated with the reference state.}

Having this additional structure on $G$, we assign the length of a trajectory $\gamma \colon \mathbb{R}\rightarrow G$ between two endpoints $\gamma(0)$ and $\gamma(1)$, with its associated tangent vector $\dd\gamma/\dd t \coloneqq \dot{\gamma}(t)$, as
\begin{equation}
    \ell_\gamma = \int_0^1 \dd t \sqrt{g_{\gamma(t)}(\dot{\gamma}, \dot{\gamma})} = \int_0^1 \dd t\, \norm{\dot{\gamma}(t)}.
\end{equation}
The complexity of some element $M\in G$ is then simply defined as the geodesic length between the identity and $M$,
\begin{equation}\label{eq:RiemannComplexityUnitary}
    \mathcal{C}(M) = \min_{\gamma}\{\ell_\gamma\,\,| \,\,\gamma(0)=\mathds{1}, \gamma(1) = M\}.
\end{equation}
This associates a value of `complexity' to a given Gaussian transformation. 

However, this is not enough to talk about the complexity of a given Gaussian state. To define state complexity, a second minimization is necessary, since there is more than one Gaussian transformation that connects the reference and target states.
Recall that for bosons, general Gaussian transformations take the form $M = (\bm{v}, S)$ where $\bm{v} \in \mathbb{R}^{2N}$ and $S$ is a symplectic matrix. Now, given a bosonic Gaussian state $\ket{J_R, \bm{0}}\equiv \ket{J_R}$ as a reference, and taking another bosonic Gaussian state $\ket{J_T, \bm{z}_T}$ as a target (possibly with nonzero displacement), the state complexity of $\ket{J_T, \bm{z}_T}$ relative to $\ket{J_R}$ is defined as
\begin{equation}\label{eq:RiemannComplexityStateBoson}
    \mathcal{C}(\ket{J_R}, \ket{J_T, \bm{z}_T}) = \min_{S}\{\mathcal{C}(\bm{z}_T, S)\,\,|\,\, J_T = SJ_RS^{-1}\},
\end{equation}
where $S$ ranges over the symplectic group. Note that the displacement part of the Gaussian transformation is univocally determined by the vector of means of the target state, but the symplectic transformation is only fixed up to the action of an element of the stabilizer of $J_R$---which is why the minimization over $S$ above is required. For fermions, we do not have to consider displacements, and the state complexity of a given target state $\ket{J_T}$ relative to the reference state $\ket{J_R}$ is just given by
\begin{equation}\label{eq:RiemannComplexityStateFermion}
    \mathcal{C}(\ket{J_R}, \ket{J_T}) = \min_O\{\mathcal{C}(O)\,\,|\,\, J_T = OJ_RO^{-1}\},
\end{equation}
where $O$ is an element of the special orthogonal group, and we assume that both reference and target in this case belong to the same parity sector~\cite{Hackl2018}.  
Eqs.~\eqref{eq:RiemannComplexityStateBoson} and~\eqref{eq:RiemannComplexityStateFermion} are just the minimal geodesic length between the identity and the submanifold given by the equivalence class of Gaussian transformations that prepare the target state taking  the reference state as the starting point. In all cases, the differential equations imposing the extremization of the length between reference and target~\eqref{eq:RiemannComplexityUnitary}-\eqref{eq:RiemannComplexityStateFermion} in any given coordinate chart $\gamma^i$ for the group manifold $G$ is simply the good old geodesic equations,
\begin{equation}\label{eq:geodesicequation}
    \dfrac{\dd^2 \gamma^i}{\dd s^2} + \underbrace{\dfrac{1}{2}g^{il}\left(\partial_j g_{kl} + \partial_k g_{lj}-\partial_l g_{jk}\right)}_{\equiv \Gamma^i_{\phantom{i}jk}}\dfrac{\dd \gamma^j}{\dd s}\dfrac{\dd \gamma^k}{\dd s} = 0,
\end{equation}
where $s$ is an affine parameter along the curve $\gamma(s)$ (that is, a parameter such that the norm of the tangent vector to the trajectory is kept constant), and $\Gamma^{i}_{\phantom{i}jk}$ are the standard Christoffel symbols. 

Given a metric tensor on the group manifold $G$, we can define the \emph{Riemannian exponential map} from tangent vectors at the identity (which we identify with elements of the Lie algebra $\mathfrak{g}$ of $G$) to points on the manifold, which we will denote by $\exp$. More specifically, the Riemannian exponential is a function that assigns to every tangent vector $V\in T_{\mathds{1}}G \simeq \mathfrak{g}$ a group element $\exp(V) = \gamma^{(\textsc{r})}_V(1)$, where $\gamma^{(\textsc{r})}_V(t)$ is the geodesic that starts from $\gamma^{(\textsc{r})}_V(0) = \mathds{1}$ with tangent vector $\dot{\gamma}^{(\textsc{r})}_V(t)|_{t=0} = V$. Equivalently, $\exp(V)$ corresponds to the endpoint of a geodesic that starts at $\mathds{1}$ with tangent vector $V$ and has total length $\norm{V}$. 

The fact that $G$ is a Lie group also equips it with another notion of exponentiation, namely the \emph{Lie exponential map}. For a given element $V$ of the Lie algebra, we define the Lie exponential of $V$ as the point $e^V = \gamma^{(\textsc{l})}_V(1)$, where $\gamma^{(\textsc{l})}_V(t)$ is the flow line of the right-invariant vector field on $G$ whose value at the identity is $\dot{\gamma}^{(\textsc{l})}_V(t)|_{t=0} = V$. 

Note that these exponentials (the Riemannian exponential here denoted by $\exp(V)$, and the Lie exponential denoted by $e^V$) are two distinct maps. The Lie exponential is a purely group-theoretical concept which does not depend on a metric at all, whereas the Riemannian exponential is only defined once the manifold is equipped with a metric. For a completely general Riemannian metric, the Riemannian and Lie exponentials will not coincide; in fact, they are only guaranteed to be equal if the metric is bi-invariant (that is, both left- and right-invariant) under the group action. In the case of fermions this can be achieved by taking the metric to be the negative of the Killing form on the group manifold $\text{SO}(2N)$. For bosons, on the other hand, this is not an option, since the Killing form on $\Sp$ does not have a definite sign, so it cannot be used to define a Riemannian metric.

For our purposes, however, we actually do \emph{not} want both exponentials to match in general. Even for the case of fermions (where a bi-invariant metric can be defined), we would still like to accommodate for a complexity measure that is not right-invariant, due to the considerations made at the end of Section~\ref{Sec:ComplexityBackground}. Instead, all that we will demand is that the two maps coincide only for some properly chosen subgroup of the Lie algebra $\mathfrak{g}$, which we now describe. 

Consider the tangent space to the identity of $G$, which is identified with the Lie algebra $\mathfrak{g}$. Pick a reference state characterized by a complex structure $J_{R}$, which, as explained in Sec.~\ref{Sec:GaussianStatesParametrization}, selects a stabilizer $\text{Sta}_{J_R}(N) \subset G$. From now on, to avoid too much clutter in the notation, we will omit the subscript $J_R$ from the stabilizer, and just keep in mind that all statements that depend in some way on the stabilizer will implicitly depend on the choice of reference state. The stabilizer subgroup $\text{Sta}(N)$, being a Lie group of its own, also possesses a Lie algebra, which we will denote by $\mathfrak{sta}(N) \subset \mathfrak{g}$. 
For the purposes of defining a geometric measure of complexity, it is very natural to at least expect that the image of both exponential maps is the same when acting on $\mathfrak{sta}(N)$. After all, if this were not the case, there would be elements of the stabilizer whose lowest-cost preparation included Gaussian transformations that are not part of the stabilizer---which does not make a lot of sense, since we would like to have circuits that act nontrivially on the reference state to have strictly larger costs than those which do not.

We are now in a position to establish the one condition that we will assume the metric on $G$ should satisfy:
\begin{enumerate}
    \item The metric must be such that, for every $V\in \mathfrak{sta}(N)$, the trajectory $M(t) = e^{tV}$ is a geodesic, and the parallel transport along that geodesic of an arbitrary vector from $\mathds{1}$ to $M(t)$ is given by right-translation by $M(t)$.
\end{enumerate}
This condition is more restrictive than just requiring that both the Riemannian and Lie exponentials span the same submanifold when acting on $\mathfrak{sta}(N)$. For one, we are prescribing how parallel transport acts also on elements of $\mathfrak{g}$, that do not belong to $\mathfrak{sta}(N)$. Moreover, we are assuming that for every $V$ in the Lie algebra of the stabilizer, $\exp(V) = e^V$---in other words, the images of both maps acting on $\mathfrak{sta}(N)$ are the same element-by-element. We nevertheless consider these constraints to be very natural: the stabilizer group ultimately corresponds to the ``trivial'' directions on the space of Gaussian operations from the point of view of state complexity. Therefore, imposing these constraints on the metric for the stabilizer is not a strong restriction, and it will still allow us to define a geometric measure of complexity in full power. Namely, we note that we still have enough degrees of freedom to meaningfully choose a complexity metric for a given general setup, since we have not specified the parallel transport along geodesics that are not contained in the stabilizer. This is precisely where we have the freedom to apply physical constraints on the available resources in order to build the complexity geometry.

Equipped with the metric at the identity, we can split $\mathfrak{g}$ as a direct sum $\mathfrak{sta}(N)\oplus \mathfrak{sta}_{\perp}(N)$, where $\mathfrak{sta}_{\perp}(N)$ is the orthogonal complement of $\mathfrak{sta}(N)$,
\begin{equation}
     \mathfrak{sta}_{\perp}(N) = \{K \in \mathfrak{g}, g_{\mathds{1}}(K, V) = 0 \,\,\forall\, V \in \mathfrak{sta}(N)\}.
\end{equation} 
We can then locally foliate $G$ near the identity element in terms of the cylinder-like surfaces defined by
\begin{equation}\label{eq:defradius}
    C_{r} = \{\exp(V)S\,\,|\,\, V\in \mathfrak{sta}_{\perp}(N), \norm{V} = r, S\in\text{Sta}(N)\}.
\end{equation}
Recall that due to the assumption previously made, $\exp(tV)S$ is precisely the geodesic starting from $S$ at $t=0$ with tangent vector equal to the parallel transport of $V$ from the identity to $S$---which is orthogonal to $\text{Sta}(N)$ and has the same norm as $V$ because parallel transport preserves inner products. Therefore, in plain English, $C_r$ is a codimension-$1$ surface generated by taking the submanifold $\text{Sta}(N)$, and then shooting off geodesics of length $r$ in the orthogonal directions to $\text{Sta}(N)$.

\begin{figure*}
    \includegraphics[width=\textwidth]{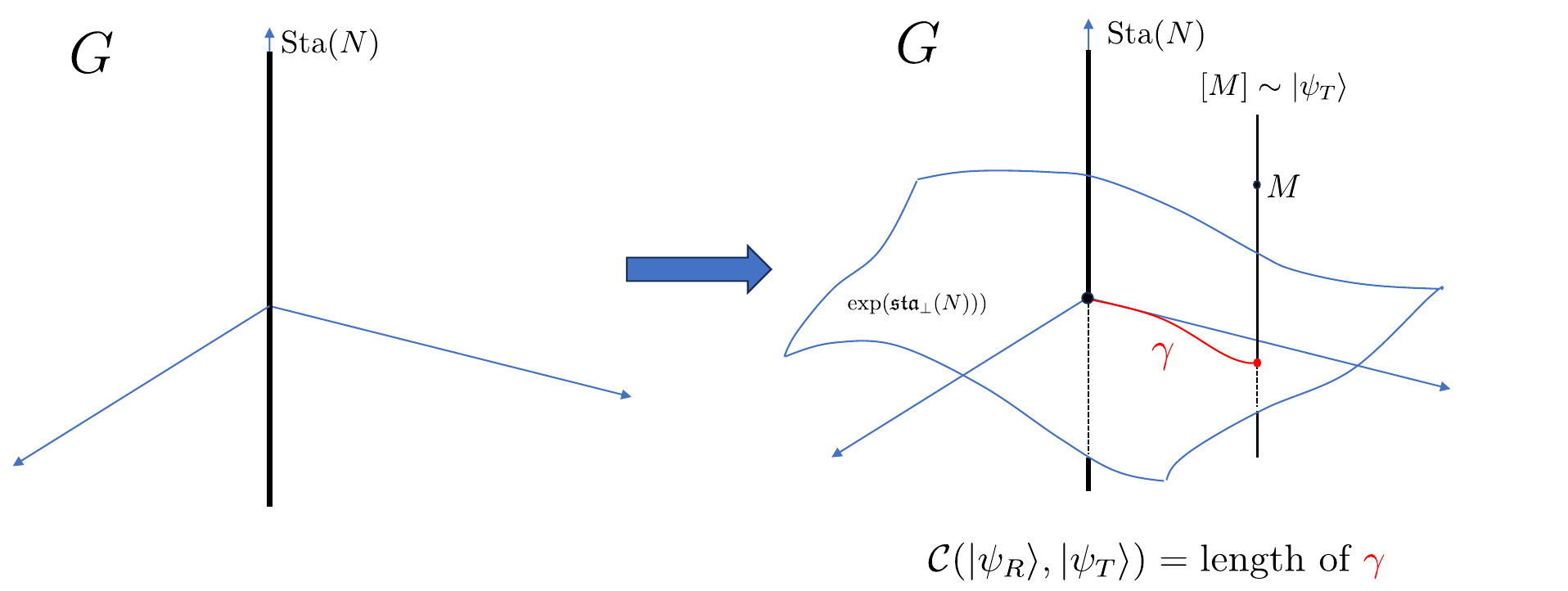}
    \caption{Pictorial representation of the geometrical ingredients involved in defining state complexity from a metric in the group manifold. After fixing the reference state (and thus the stabilizer subgroup $\text{Sta}(N)$ in $G$), one shoots off geodesics in the orthogonal directions to the generators of the stabilizer at the identity, thus constructing the submanifold $\exp(\mathfrak{sta}_{\perp}(N))$. The state complexity of the target state $\ket{\psi_T}$ given by the equivalence class of some Gaussian transformation $M\in G$ is then visualized as the length of the geodesic connecting $\mathds{1}$ to the point where the equivalence class $[M]$ intersects $\exp(\mathfrak{sta}_{\perp}(N))$.}
    \label{fig:complexityvisualization}
\end{figure*}

This setup allows us to talk about the state complexity of a given state relative to a reference state in terms of the radial coordinate $r$ defined in~\eqref{eq:defradius} as follows. Note that the map $(V, S) \mapsto \exp(V)S$ provides a local coordinate chart from some local neighborhood of the zero vector\footnote{Whether or not the Riemannian exponential map defines a coordinate chart in a given neighborhood of the identity of $G$ 
depends on whether the Riemann curvature causes caustics or not. We will make the reasonable assumption that caustics are not present within the range of target states we wish to consider.} in $\mathfrak{sta}_{\perp}(N)$ (times the stabilizer group) to some local neighborhood of the identity element on the group manifold $G$. By the way the coordinate chart is constructed, two points in the same equivalence class by the action of the stabilizer have the same $V$ coordinates; in particular, they belong to the same cylinder $C_r$. But the distance of every point of the cylinder to the identity is lower-bounded by the cylinder's radius, and there always exist trajectories whose length is precisely equal to that radius---namely, those that emanate from the identity in the orthogonal direction to $\mathfrak{sta}(N)$ and whose tangent vector has magnitude equal to $r$. We have therefore shown that the radial coordinate \emph{is} the state complexity. A geometric visualization of the strategy is displayed in Fig.~\ref{fig:complexityvisualization}.

This construction draws on and extends the strategy introduced in~\cite{Hackl2018, Chapman2019} to a much wider range of Riemannian metrics defining the complexity geometry of Gaussian states. In particular, it highlights that the key feature which made the proof of their complexity bound possible---namely, the fact that every Gaussian transformation on the same equivalence class belonged to the same cylinder $C_r$---did not actually depend on the assumption of right-invariance or on the definition of the inner product at the identity. This observation is ultimately what allowed us to leverage that construction and generalize the characterization of state complexity to more general (and potentially more physically motivated) complexity metrics.

\section{A non-reversible measure of complexity}\label{sec:notimereversal}

A common feature in the examples of geometric measures of complexity found in the literature as well as those mentioned so far in this paper is that they are all ``time-reversal-symmetric''---i.e., the complexity is invariant if the roles of reference and target states are reversed\footnote{Strictly speaking this is not exactly true in the example of coherent states that will be considered in Subsec.~\ref{sub:coherentstates}, since we assumed by convention that the reference state would always be chosen to have vanishing vector of means. In that case, reverting the roles of reference and target states must also be accompanied by a trivial overall displacement that centers the new reference state back at the origin, which results in the modification from $(\ket{J_R}, \ket{J_T, \bm{z}_T})$ to $(\ket{J_T}, \ket{J_R, -\bm{z}_T})$ as reference and target. It is clear that the state complexity~\eqref{eq:coherentstatecomplexity} is invariant under this replacement as well.}. This is a simple consequence of the fact that all cost functions presented thus far assign the same cost to the direction $V$ and to the time-reversed direction $-V$, for every tangent vector $V$ at any given point in the group manifold $G$. 

From an experimentalist's point of view, however, it is natural to imagine contexts in which the ``difficulty'' in connecting two given states is \emph{not} symmetric under swapping the roles of reference and target states. This may happen due to some effectively irreversible interaction with an external environment which drives the system to some preferred state: if there is some decay mechanism by which the system tends to spontaneously flow to the ground state of its free Hamiltonian, for instance, it is reasonable to conceive of a complexity measure which would naturally assign a lower cost to the circuit that starts at an excited state and ends at the aforementioned ground state than to the circuit that goes the other way around. It is therefore interesting to ask what new structures could be added to our geometrical measure of complexity, in order to encode some form of time reversal symmetry breaking.

\subsection{Formalism for pure states}\label{sub:IrreversiblePureStates}

Although loss of time reversal symmetry is often tied to interactions with an uncontrollable environment that makes the dynamics irreversible, experimental asymmetries in the generation of different states appears even in cases where we model state preparation with unitary dynamics. It should therefore be possible to incorporate this feature in models which retain unitarity from the perspective of the system of interest. A simple complexity measure that readily achieves this is given by the following ``generalized length'' functional to the curve $\gamma(t)$ on the group manifold:
\begin{equation}\label{eq:EMcomplexity}
    C_{\gamma} = \int_0^1 \dd t\,\norm{\dot{\gamma}(t)} - \int_0^1 \dd t \,A_i(\gamma(t))\dot{\gamma}^i.
\end{equation}
In addition to the metric $g$, this cost functional also depends on a background $1$-form $A_i$. Since the metric is assumed to be non-degenerate, this $1$-form can equivalently be seen as arising from a background vector field $A^i \equiv g^{ij}A_j$, so that~\eqref{eq:EMcomplexity} becomes
\begin{equation}\label{eq:EMcomplexity2}
    C_{\gamma} = \int_0^1 \dd t\,\norm{\dot{\gamma}(t)} - \int_0^1 \dd t \,g_{\gamma(t)}(A, \dot{\gamma}).
\end{equation}
In order for the cost function implied by Eq.~\eqref{eq:EMcomplexity2} to be positive-definite, it suffices for the norm of the background vector field $A^i$ according to the metric $g$ to be smaller than or equal to $1$. This is because the Cauchy-Schwarz inequality guarantees that
\begin{equation}
    |g(A, \dot{\gamma})| \leq \norm{A}\,\norm{\dot{\gamma}},
\end{equation}
and therefore demanding $\norm{A} \leq 1$ everywhere in the group manifold is enough to ensure that
\begin{equation}
    \norm{\dot{\gamma}} - g(A, \dot{\gamma}) \geq 0,
\end{equation}
which in turn guarantees that the cost function in Eq.~\eqref{eq:EMcomplexity} is always nonnegative. It is also clear that this modified cost function is smooth, satisfies the triangle inequality, and is positive-homogeneous in $\dot{\gamma}$, but it is \emph{not} invariant under time reversal since the costs of going in the direction $\dot{\gamma}$ or its reverse $-\dot{\gamma}$ now differ by the inner product $-2A_i \dot{\gamma}^i$. Note that in the limiting case where $\norm{A}=1$ at a given point $p$ in the group manifold, there is one nontrivial direction at that $p$ with zero cost---namely, the trajectory with tangent vector parallel to $A^i$ at $p$. 

One can immediately recognize the complexity functional~\eqref{eq:EMcomplexity} as mathematically identical to the action of a charged particle moving in a background electromagnetic field, with the $1$-form $A_i$ being analogous to the vector potential. The equation of motion for the trajectory on circuit space that extremizes~\eqref{eq:EMcomplexity} thus takes the same form as the Lorentz force law in a generally curved background metric,
\begin{equation}\label{eq:LorentzForce}
    \dfrac{\dd^2 \gamma^i}{\dd s^2} + \Gamma^{i}_{\phantom{i}jk}\dfrac{\dd \gamma^j}{\dd s}\dfrac{\dd \gamma^k}{\dd s} = F^{i}_{\phantom{i}j}\dfrac{\dd \gamma^j}{\dd s},
\end{equation}
where $\Gamma^{i}_{\phantom{i}jk}$ is given by the same expression as in~\eqref{eq:geodesicequation}, and we have $F^{i}_{\phantom{i}j} \equiv g^{ik}F_{kj}$, with $F_{ij}$ the usual standard electromagnetic field strength,
\begin{equation}
    F_{ij} = \partial_i A_j - \partial_j A_i.
\end{equation}
\subsection{Generalizations to mixed states}

The formalism from Sec.~\ref{sub:IrreversiblePureStates} can be readily adapted to also include mixed states. The usual strategy to define complexity for mixed states is to consider the so-called \emph{purification complexity}~\cite{Agon2019}. Given a Hilbert space $\mathcal{H}$ describing the system of interest, for any two density matrices $\hat{\rho}_R$ and $\hat{\rho}_T$ corresponding to the reference and target states, the purification complexity of $\hat{\rho}_T$ relative to $\hat{\rho}_R$ is given by
\begin{equation}\label{eq:purificationcomplexity}
    \mathcal{C}(\hat{\rho}_R, \hat{\rho}_T) = \min_{\ket{\psi_R}, \ket{\psi_T}}\Big\{\mathcal{C}(\ket{\psi_R}, \ket{\psi_T})\Big\}
\end{equation}
where the minimization ranges over states $\ket{\psi_R}, \ket{\psi_T}$ in an enlarged Hilbert space $\mathcal{H}\otimes\mathcal{H}'$ such that 
\begin{align}
    \hat{\rho}_R = \Tr\big(\ket{\psi_R}\bra{\psi_R}\big)_{\mathcal{H}'}, \\
    \hat{\rho}_T = \Tr\big(\ket{\psi_T}\bra{\psi_T}\big)_{\mathcal{H}'}.
\end{align}
If both the reference and target states are Gaussian, it is common to further restrict the minimization above to only consider purifications that are also Gaussian~\cite{Caceres2020, Camargo2021}. In this case, one might call the complexity measure given by Eq.~\eqref{eq:purificationcomplexity} the \emph{Gaussian} complexity of purification. It is clear that, once we have a generalized cost functional that displays some feature of non-reversibility for pure states---given, for instance, by the additional background structure described in Sec.~\ref{sub:IrreversiblePureStates}---, the same feature will be generally transported to the case of mixed states according to the definition above. A similar logic also applies for alternative approaches to the geometrization of complexity based on directly defining a notion of distance on the space of density matrices via the Bures metric~\cite{Ruan2021}, which generalizes the approach based on the Fubini-Study metric~\cite{Pastawski2017} to the case of mixed states. 
\section{Examples}

We will consider here four examples. For ilustration purposes, the first one reviews an example already analyzed in the literature~\cite{Hackl2018, Chapman2019}. The second example extends the first, by including the case of bosonic coherent states. This has also been analyzed in the past through slightly different techniques~\cite{CoherentStateComplexity}, but here we will be able to provide an easily computable closed expression for the geometric Gaussian complexity of general coherent states based on minimal assumptions. Finally, the last two examples consider simple extensions of the geometric complexity formalism that may capture features of the difficulty of physical state preparation in a lab that right-invariant metrics cannot.

\label{Sec:Applications}
\subsection{Right-invariant metric on the group manifold}\label{sub:previousmetric}
The first example we will present serves the purpose of illustrating how the general machinery laid out in previous sections is enough to provide closed-form expressions for circuit complexity. This example will closely follow
the work in~\cite{Hackl2018, Chapman2019}, which was the main inspiration for the generalization we presented in Sec.~\ref{Sec:gengaussianmetric}.

In~\cite{Hackl2018, Chapman2019}, the authors considered pure Gaussian states with vanishing vector of means, and therefore the group $G$ for bosons is reduced to just $\Sp$. They then apply this generalized cylindrical foliation to evaluate the state complexity derived from a Riemannian metric which, at the Lie algebra $\mathfrak{g} \simeq T_{\mathds{1}}G$ (in this case equal to $\mathfrak{so}(2N)$ for fermions, and $\mathfrak{sp}(2N, \mathbb{R})$ for bosons), is given by
\begin{equation}\label{eq:previousmetric}
    g_{\mathds{1}}(V, W) = \dfrac{1}{2}\Tr\left(V\sigma_R W^\intercal \sigma_R^{-1}\right),
\end{equation}
with $\sigma_R$ being the symmetric part of the two-point correlator of the quadratures in the chosen reference state. The metric is extended to all points $M\in G$ of the group manifold by demanding right-invariance,
\begin{equation}\label{eq:previousmetric2}
    g_{M}(V, W) = g_{\mathds{1}}(VM^{-1}, WM^{-1}).
\end{equation}
 In some sense this is a natural choice of inner product in the Lie algebra, as it does not require anything beyond the group structure and some minimal information about the reference state. For fermions, $\sigma_R$ is actually independent of the state, and fully fixed by the anticommutation relations of the quadratures. In this case, Eq.~\eqref{eq:previousmetric} is nothing but the negative of the Killing form in $\text{SO}(2N)$. For bosons, the metric unavoidably depends on the choice of reference state, as there is no canonically defined positive-definite form on $\Sp$ that only depends on the group structure of the symplectic group. 

It is important to emphasize once again that, in general, one should not expect this measure of complexity to accurately quantify actual physical limitations based on any realistic lab resources. We would expect, for instance, that an experimentally motivated measure of complexity would depend nontrivially on the reference state even in the case of fermions. For bosons, Eq.~\eqref{eq:previousmetric} includes some dependence on the reference state through its covariance matrix $\sigma_R$, but the complexity measure is still not able to account for other physically motivated features such as lack of time-reversal symmetry. However, in the absence of more detailed models for the emergence of a physically motivated complexity measure, it has been common to take~\eqref{eq:previousmetric} as a paradigmatic definition of complexity geometry in the literature, which is why we are reviewing this example first before considering any modifications. 

The choice of metric defined by Eqs.~\eqref{eq:previousmetric}-\eqref{eq:previousmetric2} allows for a very explicit verification of the general strategy laid out in Sec.~\ref{Sec:gengaussianmetric}. In particular, it is possible to write a closed-form expression for the shortest path connecting the identity to the equivalence class of Gaussian transformations that prepare a given target state, and thus obtain an analytical solution to the problem of computing Gaussian state complexity. Given the complex structures $J_R$ and $J_T$ defining the reference and target states respectively, one can always define the so-called \emph{relative complex structure}
\begin{equation}
    \Delta = J_T J_R^{-1}.
\end{equation}
Then, the geodesic that leads to the target state can be given in the space of Gaussian transformations as 
\begin{equation}\label{eq:parametrizedgeodesic}
    M(\tau) = e^{\tau\log (\Delta)/2}.
\end{equation}
One can directly verify that the generator $\log (\Delta)/2 \in \mathfrak{g}$ is in the orthogonal complement of the Lie algebra of the stabilizer of the reference state $J_R$ according to the inner product of Eq.~\eqref{eq:previousmetric}, and that \mbox{$M(1) J_R M(1)^{-1} = J_T$}. 
With this, the Gaussian state complexity of both fermionic and bosonic Gaussian systems is simply given in terms of the relative complex structure as~\cite{HacklKahler2021}
\begin{equation}\label{eq:simplestatecomplexity}
    \mathcal{C}(\ket{J_R}, \ket{J_T}) =\frac{1}{{2\sqrt{2}}} \sqrt{\Tr\left(\abs{\log(\Delta)}^2\right)}.
\end{equation}

\subsection{Complexity of coherent states}\label{sub:coherentstates}

In the case of bosons, we can generalize the example in Subsec.~\ref{sub:previousmetric} to include coherent states, where the target state contains some nonzero displacement $\bm{z}_T$. This can be motivated, for example, by the fact that many important proposals for the implementation of quantum information in continuous-variable systems are based on encoding the logical information in the displacement vector, and not in the covariance matrix~\cite{Gottesman2001, PhysRevLett.100.030503, Terhal_2020, PRXQuantum.2.020101}. Our next example will thus consist of a simple extension of the formulas of Subsec.~\ref{sub:previousmetric} to the case where the target state is a coherent state.

In parallel to what we did before, all we need to do now is to define an inner product at the tangent space of the identity, with the minimal modification of now including tangent vectors that generate displacements. This inner product is then transported to the rest of the group manifold by the analog of Eq.~\eqref{eq:previousmetric2}, with the group element now being a pair $(\bm{v}, M)\in \mathbb{R}^{2N}\rtimes \Sp$. A natural\footnote{Following the same logic as with the example in Subsec.~\ref{sub:previousmetric}, this is the simplest way to define the complexity metric in the displacement sector that does not add any extra structure.} choice of inner product at the identity can be taken as
\begin{equation}
    g_{(\bm{0}, \mathds{1})}\left((\bm{v}, V), (\bm{w}, W)\right) = \sigma^{-1}_R(\bm{v}, \bm{w}) + \dfrac{1}{2}\Tr\left(V\sigma_R W^\intercal \sigma_R^{-1}\right)
\end{equation}
where $V, W\in \mathfrak{sp}(2N, \mathbb{R})$, and
\begin{equation}
    \sigma^{-1}_R(\bm{v}, \bm{w}) = \left(\sigma^{-1}_R\right)_{ab}v^a w^b.
\end{equation}
Then, given a reference state $\ket{J_R, \bm{0}}\equiv \ket{J_R}$ and a target state $\ket{J_T, \bm{z}_T}$, the optimal circuit $(M(\tau), z(\tau))$ can be parametrized as
\begin{align}
    M(\tau) &= e^{\tau\log (\Delta)/2}, \\
    \bm{z}(\tau) &= \left(M(\tau) - \mathds{1}\right)\left(M(1) - \mathds{1}\right)^{-1} \bm{z}_T,
\end{align}
where we recall $\Delta = J_T J_R^{-1}$ is the relative complex structure from the reference to the target state. Represented as a unitary operator on the Hilbert space, the trajectory above simply corresponds to the time evolution implemented by
\begin{equation}
    \hat{U}(\tau) = e^{-\ii \hat{H}\tau},
\end{equation}
where the time-independent Hamiltonian $\hat{H}$ is defined as
\begin{align}\label{eq:geodesiccoherentstates}
    \hat{H} &= \dfrac{1}{2}F_{ab}\hat{\xi}^a\hat{\xi}^b + \alpha_a \hat{\xi}^a, \\
    F_{ab} &= \dfrac{1}{2}\Omega_{ac}\left(\log\Delta\right)^{c}_{\phantom{c}b}, \nonumber \\
    \alpha_a &= \dfrac{1}{2}\Omega_{ac}N^{c}_{\phantom{c}d}\left(\bm{z}_T\right)^d.\nonumber
\end{align}
In~\eqref{eq:geodesiccoherentstates}, $N$ is given by
\begin{equation}\label{eq:defN}
    N = \log(\Delta)(\sqrt{\Delta}-\mathds{1})^{-1}
\end{equation}
and we recall that the symplectic matrix $\Omega_{ab}$ is the inverse of  $\Omega^{ab}$ as defined in~\eqref{eq:sympmatrix}---which, for bosonic systems, is state-independent and fixed by the algebra of the quadratures. 

We thus obtain the state complexity
\begin{equation}\label{eq:coherentstatecomplexity}
    \mathcal{C}(\ket{J_R}, \ket{J_T, \bm{z}_T}) = \dfrac{1}{2}\sqrt{\dfrac{\Tr\left(\abs{\log(\Delta)}^2\right)}{2} + G(\bm{z}_T, \bm{z}_T)},
\end{equation}
where $G(\bm{z}, \bm{z}) = G_{ab}z^az^b$, and the bilinear form $G$ is defined by
\begin{equation}
    G = N^{\intercal}\sigma_R^{-1}N,
\end{equation}
with $N$ defined as in~\eqref{eq:defN}. 
The case where target and reference states share the same complex structure and only differ by the displacement $\bm{z}_T$ is readily accommodated as the limit when $\Delta= \mathds{1}$. In this case, we have $N = 2\cdot\mathds{1}$, the optimal path on the space of displacements reduces to a straight line,
\begin{equation}
    \bm{z}(\tau) = \tau \bm{z}_T,
\end{equation}
and the state complexity reduces to
\begin{equation}
    \mathcal{C}(\ket{J_R}, \ket{J_R, \bm{z}_T}) = \sqrt{\sigma_R^{-1}(\bm{z}_T, \bm{z}_T)}.
\end{equation}
\subsection{Weyl transformation of right-invariant metric}\label{sub:Weyl}
It would also be interesting to have an example of complexity metric which does not have as much symmetry as the metrics presented in Subsecs.~\ref{sub:previousmetric} or~\ref{sub:coherentstates}. After all, it is reasonable to expect that a measure of complexity that is motivated by constraints from physical limitations will not be represented by a right-invariant metric, but will depend more nontrivially on the state on which a given elementary gate set is acting.

A very simple example of a metric that is no longer right-invariant, but for which most of the existing closed-form expressions of Subsec.~\ref{sub:previousmetric} will still be greatly useful, is obtained if we perform a Weyl transformation on the previously right-invariant metric. That is, we define a new metric in the group manifold in terms of the metric presented in Sec.~\ref{sub:previousmetric} by
\begin{equation}\label{eq:weyltransformed}
    \tilde{g}_M(V, W) = e^{2\omega(r)}g_{M}(V, W).
\end{equation}
In the Weyl transformation above, we are assuming that  $\omega(r)$ is a function only of the generalized radial coordinate $r$---which, in this case, for any Gaussian transformation in the equivalence class labeled by a given complex structure $J$, is just the state complexity given by Eq.~\eqref{eq:simplestatecomplexity} with $J_T = J$. Heuristically, the metric~\eqref{eq:weyltransformed} describes a simple model where the cost of acting with a certain gate may change depending on how complex the circuit already is at the time the gate is applied. 

Mathematically, the importance of $\omega(r)$ being only a function of $r$ stems from the observation that, in this case, the path given in Eq.~\eqref{eq:parametrizedgeodesic} is also a geodesic of the modified metric. To see this, first note that the geodesic equation~\eqref{eq:geodesicequation} for a curve with tangent vector $V$ can be written in more compact form as
\begin{equation}\label{eq:geodesicequation2}
    V^i\nabla_i V^j = 0,
\end{equation}
where $\nabla_i$ above denotes the Levi-Civita connection associated to $g_{ij}$---i.e., the covariant derivative that is torsion-free and compatible with the metric $g_{ij}$. 
Under a general Weyl transformation $g_{ij} \mapsto \tilde{g}_{ij} = e^{2\omega}g_{ij}$, the Christoffel symbols transform as
\begin{equation}
    \Gamma^{i}_{\phantom{i}jk} \mapsto \tilde{\Gamma}^{i}_{\phantom{i}jk} = \Gamma^{i}_{\phantom{i}jk} + \delta^i_j\nabla_k\omega + \delta^i_k\nabla_j\omega - g_{jk}g^{il}\nabla_l\omega.
\end{equation}
Therefore, if the trajectory $M(\tau)$ with tangent vector $V = \dd M/\dd\tau$ was a geodesic of the metric $g_{ij}$, the same trajectory (with the same parametrization), for the case of  the Weyl-transformed metric, will now satisfy
\begin{equation}\label{eq:WeylTransformedGeodesicEqn}
    V^i \tilde{\nabla}_i V^j = 2(V^i\nabla_i\omega)V^j - \norm{V}^2_g \,g^{ij}\nabla_i\omega,
\end{equation}
where $\tilde{\nabla}_i$ is the Levi-Civita connection associated to the new metric $\tilde{g}_{ij}$, and $\norm{V}^2_g$ corresponds to the squared norm of $V$ according to the \emph{old} metric, $\norm{V}^2_g = g_{ij}V^iV^j.$

Eq.~\eqref{eq:WeylTransformedGeodesicEqn} certainly does not look like the geodesic equation for arbitrary $\omega$ and for a general geodesic of the old metric. However, if we take the geodesic~\eqref{eq:parametrizedgeodesic} and we assume that $\omega = \omega(r)$, the problem simplifies greatly, since in this case we have that, along the trajectory~\eqref{eq:parametrizedgeodesic}, it holds that $g^{ij}\nabla_j \omega \propto V^i$.
This is because the tangent vector $V$ satisfies $V \propto \partial_r$, the conformal factor $\omega$ is such that $\dd\omega \propto \dd r$, and the vector field $\partial_r$ along the trajectory~\eqref{eq:parametrizedgeodesic} is orthogonal to the cylinders of constant $r$ (see~\cite{Hackl2018, Chapman2019}), so $g^{ij}\nabla_j\omega \propto (\partial_r)^i$. We can thus simply write
\begin{equation}\label{eq:proportionalitydwV}
    \nabla_i\omega = \pm\dfrac{\norm{\dd\omega}_g}{\norm{V}_g}\, V_i,
\end{equation}
where $\norm{\dd \omega}_{g}$ corresponds to the norm of the $1$-form $\dd\omega$ according to the metric $g$, $\norm{\dd\omega}^2_g = g^{ij}(\nabla_i\omega)(\nabla_j\omega)$. 
The plus-or-minus choice in~\eqref{eq:proportionalitydwV} just accommodates the cases where $\omega$ increases or decreases with increasing $r$, with the plus sign applying to the former and the minus sign applying to the latter. In both cases, substituting this back into~\eqref{eq:WeylTransformedGeodesicEqn}, we obtain
\begin{align}\label{eq:nonaffinelyparametrizedgeodesic}
     V^i \tilde{\nabla}_i V^j &= (V^i\nabla_i\omega) V^j \nonumber \\
     &\equiv \Omega(\tau) V^j,
\end{align}
which is of the form of a \emph{non}-affinely parametrized geodesic curve, $V^i \tilde{\nabla}_i V^j \propto V^j$, with the proportionality factor $\Omega(\tau)$ being given by $\Omega(\tau) = V^i\nabla_i\omega= \dd\omega/\dd\tau$. In the last equation, $\omega(\tau)$ should be understood as the pullback of $\omega(r)$ to the trajectory~\eqref{eq:parametrizedgeodesic}. A new affine parameter $s$ (such that the associated tangent vector $\tilde{V} = \dd M/\dd s$ satisfies the geodesic equation in the form~\eqref{eq:geodesicequation2} for the new metric $\tilde{g}_{ij}$) is then obtained by noting that, if both $\tilde{V}$ and $V$ are tangent vectors for the same curve, we can write
\begin{equation}
    V = \dfrac{\dd M}{\dd\tau} = \dfrac{\dd s}{\dd \tau} \dfrac{\dd M}{\dd s} = \dfrac{\dd s}{\dd \tau}\tilde{V}.
\end{equation}
Plugging this back into Eq.~\eqref{eq:nonaffinelyparametrizedgeodesic} and demanding that $\tilde{V}^i\tilde{\nabla}_i\tilde{V}^j = 0$ then gives us a differential equation for $s$,
\begin{equation}\label{eq:properlengthreparametrization}
    \dfrac{1}{f}\dfrac{\dd f}{\dd \tau} = \dfrac{\dd\omega}{\dd\tau}, \,\,\,\, f(\tau) = \dfrac{\dd s}{\dd \tau}.
\end{equation}
We obtain a unique solution to~\eqref{eq:properlengthreparametrization} by demanding that $s(\tau=0)=0$ and $s(\tau=1) = 1$, so that the new affine parameter also ranges from $0$ to $1$ between the reference and target states. This gives us
\begin{equation}
    s(\tau) = \dfrac{1}{\int_0^1 \dd \tau' e^{\omega(\tau')}}\int_0^\tau \dd \tau' e^{\omega(\tau')}.
\end{equation}
Putting all of this together with the results reviewed in Sec.~\ref{sub:previousmetric}, we conclude that the state complexity derived from the metric given by~\eqref{eq:weyltransformed} is simply
\begin{equation}\label{eq:weyltransfComplexity}
    \Tilde{\mathcal{C}}(\ket{J_R}, \ket{J_T}) = \dfrac{1}{2\sqrt{2}}\int_0^1\dd\tau\, e^{\omega(\tau)}\sqrt{\Tr\left(\abs{\log(\Delta)}^2\right)}
\end{equation}
In fact, once we knew that the original geodesics were purely radial and the conformal factor was only a function of $r$, there was an elementary argument that would also have led to Eq.~\eqref{eq:weyltransfComplexity}. The new line element in the radial direction after the Weyl transformation is $\dd\ell = e^{\omega(r)}\dd r$, which once integrated from $r'=0$ to $r'=r$ gives us the new proper length
\begin{equation}\label{eq:easyargument}
    \ell = \int_0^{r}\dd r'\,e^{\omega(r')} = r\int_0^{1}\dd \tau\,e^{\omega(\tau)},
\end{equation}
where we have renormalized the limits of integration by setting $\tau = r'/r$, and used the slight abuse of notation $\omega(r(\tau)) = \omega(\tau)$. By then remembering that the radial coordinate associated to the target state is precisely given by~\eqref{eq:simplestatecomplexity}, the result from Eq.~\eqref{eq:easyargument} is exactly what we obtained in Eq.~\eqref{eq:weyltransfComplexity}.
\subsection{Non-reversible complexity for a single\\ bosonic mode}\label{sub:singlemodecomplexity}
Finally, as a simple example to illustrate a setting that also includes non-reversibility, let us consider a single bosonic degree of freedom, and restrict ourselves to the subset of pure Gaussian states with vanishing vector of means. In this case, the full space of pure Gaussian states under consideration can be characterized by $2\times 2$ covariance matrices $\sigma$. Without loss of generality, we can set the covariance matrix of the reference state to just be the identity,
\begin{equation}
    \sigma_R = \begin{pmatrix}
        1 & 0 \\
        0 & 1
    \end{pmatrix}.
\end{equation}
We then recall that the symplectic matrix for bosons is state-independent and given by
\begin{equation}
    \Omega = \begin{pmatrix}
        0 & 1 \\
        -1 & 0
    \end{pmatrix},
\end{equation}
so the reference state can equivalently be characterized by a complex structure which, in the simple choice of canonical basis where $\sigma_R$ becomes the identity, is just represented as
\begin{equation}\label{eq:refcomplexstructure}
    J_R = \begin{pmatrix}
        0 & 1 \\
        -1 & 0
    \end{pmatrix}.
\end{equation}
Thanks to the simplicity of the setup, all of the general ingredients developed in the previous sections can be readily worked out in full detail. The stabilizer subgroup $\text{Sta}(J_R)$ associated to the complex structure~\eqref{eq:refcomplexstructure} is simply the orthogonal group in $2$ dimensions, whose Lie algebra $\mathfrak{sta}(J_R) \simeq \mathfrak{so}(2)$ is one-dimensional and generated by the single element $\ii\sigma_{y}$, where $\sigma_y$ is the second Pauli matrix. The orthogonal complement to $\mathfrak{sta}(J_R)$ within $\mathfrak{sp}(2, \mathbb{R})$ according to the standard choice of inner product~\eqref{eq:previousmetric} can be parametrized as
\begin{equation}
    \mathfrak{sta}_\perp(J_R) = \left\{r\begin{pmatrix}
        \cos\phi & \sin\phi \\
        \sin\phi & -\cos\phi
    \end{pmatrix} \,\,\Big|\,\, 0\leq \phi<2\pi, r\geq 0 \right\}.
\end{equation}
The exponentiated version of these elements of the Lie algebra then provides us with elements of the symplectic group which can be recognized as general single-mode squeezings $S(r, \phi)$, where $\phi$ determines the orientation of the squeezing, and $r$ quantifies its magnitude. The parameters $r$ and $\phi$ serve as coordinates on the space of pure Gaussian states of a single bosonic mode with vanishing vector of means. The fact that the squeezing magnitude is notated as $r$ is particularly convenient, as it is simple to verify---with the complexity geometry defined by  the metric of Subsec.~\ref{sub:previousmetric}---that $r$ corresponds precisely to the complexity of the state obtained by acting on the complex structure of the reference state with the squeezing operator $S(r, \phi)$ as $J = S(r, \phi)J_RS(r, \phi)^{-1}$. In other words, the squeezing magnitude $r$ is also the radial coordinate characterizing state complexity according to the right-invariant metric defined by Eqs.~\eqref{eq:previousmetric}-\eqref{eq:previousmetric2} in the case of a single bosonic degree of freedom. That same metric restricted to the $(r, \phi)$-plane gives us the line element
\begin{equation}
    \dd s^2 = \dd r^2 + \cosh(2r)\sinh^2(r)\,\dd\phi^2.
\end{equation}
In this case, a very natural addition that reflects some level of non-reversibility would be, for instance, to assign directions of increasing $r$ a higher cost than those of decreasing $r$. As suggested in Sec.~\ref{sec:notimereversal}, this can be achieved by simply adding to the cost function the contribution from a vector potential of the form
\begin{equation}
    A = -f(r, \phi)\,\dd r,
\end{equation}
for some positive function $f(r, \phi)$. In the simplest case where $f$ is only a function of $r$, the vector potential is just a total derivative, and can be written as $A = -\dd h$ for some $h(r)$ such that $f = \dd h/\dd r$. A vector potential of this form does not change what the optimal path is, since in this case $F_{ij}=0$ and therefore the equation of motion~\eqref{eq:LorentzForce} is the same as if there were no vector potential at all. It does however change the actual value of the complexity~\eqref{eq:EMcomplexity} by an additional term of \mbox{$\Delta h = h(r_f)-h(0)$}, with $r_f$ being the squeezing parameter of the final state relative to the reference state. If the roles of reference and target are reversed, the contribution from this term changes to $-\Delta h$, and therefore some notion of non-reversibility is already present. If $f$ also depends on $\phi$, then on top of the asymmetry between reference and target states, we also get some nontrivial modification to the equation of motion~\eqref{eq:LorentzForce}, analogous to an external magnetic field orthogonal to the $(r, \phi)$-plane.

As a final note, we remark that including a conformal factor of the form discussed in Subsec.~\ref{sub:Weyl} in this case has a very intuitive interpretation as well: namely, it means that applying the same gate to the state of the system becomes harder the more squeezed the state already is (something that any quantum optician may confirm since it becomes extremely difficult to achieve squeezing beyond 15 dB in quantum optics~\cite{Squeezing2, SqueezingPaper}).

\section{Conclusions}\label{ConclusionSection}

In this paper we have proposed a geometric notion of circuit complexity for Gaussian states of both bosonic and fermionic systems that has a built-in way of accounting for 1) the presence of possible inhomogeneities (which change the cost of a given gate depending on what state it acts on), and 2) the presence of non-reversibilities (which assign different costs to paths related by time reversal). This allows us to build a geometric complexity measure that can capture features of the physical complexity of state preparation that one can experience in practice. 

Instead of focusing on one particular choice of complexity measure, we kept as much generality as possible in the form of complexity functional based on Riemannian metrics on the group of Gaussian transformations. This allowed us to highlight a set of minimal assumptions that the Riemannian metric should satisfy in order to potentially be a good physical measure of how complex a given state is. With that, we were then able to generalize a previous characterization of state complexity considered in the literature~\cite{Hackl2018, Chapman2019, HacklKahler2021} thanks to a generalized cylindrical foliation of the space of Gaussian transformations that applies to any Riemannian metric satisfying the assumptions of Sec.~\ref{Sec:gengaussianmetric}. This shows how to preserve/adapt a lot of the techniques and results from~\cite{Hackl2018, Chapman2019, HacklKahler2021}, while at the same time being more flexible with the specifics of the complexity metric---which, among many other things, can naturally model the fact that in quantum optics, for instance, squeezing can be more or less costly depending on how squeezed the state already is. 

We also proposed an extension to the framework that includes an additional term in the cost functional which makes directions on the space of Gaussian transformations that are connected by time-reversal have different costs. We consider this to be a well-motivated feature that a physical measure of circuit complexity (understood as a quantifier of the ``hardness'' of preparing a given state or unitary) may display: after all, from an experimentalist's perspective, given two states $\ket{\psi_1}$ and $\ket{\psi_2}$, it is not uncommon to encounter situations where the experienced difficulty in preparing $\ket{\psi_1}$ starting from $\ket{\psi_2}$ is not the same as starting from $\ket{\psi_1}$ and then preparing $\ket{\psi_2}$. This is realized in the model proposed in Sec.~\ref{sec:notimereversal} through the addition of a background $1$-form on the space of Gaussian transformations which behaves like a background vector potential. 

Despite the fact that Gaussian dynamics is restricted to systems with quadratic Hamiltonians, the reasoning employed in this paper may also apply equally well to more general cases (potentially including interactions) as long as the generators of the elementary gate set form a Lie algebra. It would be particularly interesting to see how this plays out in the case of CFTs, especially due to its possible repercussions on considerations about holographic complexity. In particular, this general class of complexity functionals could potentially serve as a concrete boundary counterpart to the recent ``Complexity $=$ Anything'' proposal for holographic complexity in the AdS/CFT correspondence~\cite{ComplexityEqualsAnything1, ComplexityEqualsAnything2}. In this context, the infinite family of gravitational observables in asymptotically AdS spacetimes proposed as candidate duals to circuit complexity of the CFT state would be interpreted as arising from the infinite family of complexity geometries that can be assigned to the boundary theory. 

An important problem to be pursued next is to see how to provide a class of measures of complexity in quantum field theory that is explicitly derived/motivated by constraints on the physical operations (and their associated costs) that can act on the field. This was our main motivation for starting this work, and we consider that this paper establishes the first step in that direction. More specifically, we aim to understand whether considerations about the action of \emph{local probes} interacting with the field can provide some notion of complexity in quantum field theory. The use of local probes as a handle to the study of information-theoretic properties of QFT is at the heart of the field of Relativistic Quantum Information (RQI), where local probes (often called \emph{particle detectors} in this context) have been shown to provide important insights into several aspects of the interface between quantum information and quantum field theory, while also directly modelling experimentally realizable setups (see, e.g., ~\cite{Richard}). Examples of the use of particle detectors in RQI as proxies of fundamental properties of QFT include the formulation of a measurement theory for relativistic QFTs~\cite{JPGFramework}, an operational approach to entanglement in field theory~\cite{Reznik2003, Reznik1, Pozas-Kerstjens:2015, kelly} and the inference of general properties of the background spacetime from the statistics of detectors~\cite{Terno2016, TalesGeometry}, among many others. It is therefore natural to consider that similar considerations about local probes may also shed light into some notion of complexity in QFT, by providing an operationally motivated sense of complexity associated to a given family of gates that can be implemented in the field through local probes. It should be interesting to know whether the features of the dynamics between fields and probes can yield some equation of motion that privileges a subset of the metrics among those satisfying the constraints of Sec.~\ref{Sec:gengaussianmetric}, as well for the background vector potential introduced in Sec.~\ref{sec:notimereversal}.

Finally, it would also be important to understand how such motivated measures of complexity from local probes may be potentially related to other proposals for definitions of complexity from first principles. Examples include a notion of complexity defined in connection with resource theories~\cite{YungerHalpern2022}---which is also motivated by the need to accommodate constraints on what realistic agents can accomplish---and emergent concepts of complexity in the context of AdS/CFT~\cite{Caputa2017, Camargo2019, erdmenger2022complexity, Chandra2023}. We expect this to be subject of further study in the near future.

\section{Acknowledgements}
BSLT acknowledges support from the Mike and Ophelia Lazaridis Fellowship. EMM acknowledges the support of the NSERC Discovery program as well as his Ontario Early Researcher Award. Research at Perimeter Institute is supported in part by the Government of Canada through the Department of Innovation, Science and Economic Development Canada and by the Province of Ontario through the Ministry of Colleges and Universities.

\appendix

\bibliography{references}

\end{document}